\documentclass[fleqn,usenatbib]{rasti}

\usepackage{newtxtext,newtxmath}

\usepackage[T1]{fontenc}

\DeclareRobustCommand{\VAN}[3]{#2}
\let\VANthebibliography\thebibliography
\def\thebibliography{\DeclareRobustCommand{\VAN}[3]{##3}\VANthebibliography}

\usepackage{graphicx}	
\usepackage{amsmath}	
\usepackage{multirow} 

\title[Solar Jet Extraction Tool]{SJET: An Interactive Solar Jet Extraction Tool}

\author[Song Tan et al.]{
Song Tan,$^{1,2}$\thanks{E-mail: stan@aip.de}
Alexander Warmuth,$^{2}$
Frédéric Schuller,$^{2}$
Yuandeng Shen,$^{3}$
Yue Fang,$^{4}$
Jake A. J. Mitchell,$^{2}$
\newauthor
Zedong Liu$^{5}$
\\
$^{1}$Institut für Physik und Astronomie, Universität Potsdam, Karl-Liebknecht-Straße 24/25, 14476 Potsdam, Germany\\
$^{2}$Leibniz-Institut für Astrophysik Potsdam (AIP), An der Sternwarte 16, 14482 Potsdam, Germany\\
$^{3}$State Key Laboratory of Solar Activity and Space Weather, School of Aerospace, Harbin Institute of Technology, Shenzhen 518055, China\\
$^{4}$School of Physics, Anhui University, Hefei 230601, China\\
$^{5}$University of Electronic Science and Technology of China, Chengdu 610054, China
}

\date{Accepted XXX. Received YYY; in original form ZZZ}

\pubyear{\the\year{}}

\begin{document}
\label{firstpage}
\pagerange{\pageref{firstpage}--\pageref{lastpage}}
\maketitle

\begin{abstract}
Solar jets are dynamic collimated plasma flows in the solar atmosphere that play crucial roles in coronal heating and solar wind acceleration. Their complex and diverse morphologies pose significant challenges for developing universal algorithms for automatic identification and extraction, particularly for on-disk jets affected by projection effects and background contamination. We present SJET, an interactive tool for solar jet feature extraction using multiple algorithms developed in Python that integrates five thresholding algorithms with morphological operations. SJET implements a novel method for identifying start and end points based on circular regions that objectively determines jet propagation direction by exploiting morphological asymmetry, combined with modeling the axis using quadratic Bézier curves for accurate extraction of geometric parameters including length, width, curvature, and deflection angles. Validation analyses using Solar Orbiter/EUI high-resolution image and SDO/AIA observations demonstrate SJET's effectiveness across different observational conditions, with good agreement compared to traditional analysis methods, though the tool's accuracy remains dependent on user-defined threshold parameters and region of interest selection. SJET provides a solution to  method inconsistency in solar jet research through standardized processing workflows, establishing a technical foundation for large-sample statistical studies.
\end{abstract}




\section{Introduction}
\label{sec1}
Solar jets are ubiquitous dynamic phenomena in the solar atmosphere, manifested primarily as collimated plasma flows ejected from the chromosphere or transition region into the corona. Solar jets exhibit rich physical processes over multiple spatial and temporal scales \citep{2016SSRv..201....1R,2021RSPSA.47700217S,2022AdSpR..70.1580S}. From a microscopic perspective, jets represent direct manifestations of magnetic reconnection processes, with their formation involving changes in magnetic field topology and rapid energy release \citep{1994xspy.conf...29S,1995Natur.375...42Y,2021ApJ...915...17X}. From a macroscopic viewpoint, jets participate in the transport of coronal mass and energy into interplanetary space, potentially serving as important mechanisms for solar wind acceleration \citep{2013ApJ...770....6Y,2021ApJ...912L..15T,2023ApJ...942L..22D,2024ApJ...968..110D}. Recent studies have further demonstrated that small-scale jets may contribute significantly to coronal heating through collective effects \citep{2019Sci...366..890S,2023Sci...381..867C,2024A&A...686A.279S,2024ApJ...963....4S,2025ApJ...979..195D}. With advances in observational technology, particularly through the deployment of missions such as Hinode, STEREO, the Interface Region Imaging Spectrograph (IRIS, \cite{2014SoPh..289.2733D}), the Solar Dynamics Observatory (SDO, \cite{2012SoPh..275....3P}), and the Solar Orbiter \citep{2020A&A...642A...1M}, our understanding of solar jets is evolving from phenomenological descriptions toward physical mechanism analysis. However, the complex morphological features revealed by high-resolution observational data pose new challenges for solar jet analysis \citep{2017ApJ...851...67S,2021ApJ...913...59W,2024A&A...687A.172J,2024A&A...691A.198J,2025A&A...702A..88T,2025A&A...702A.189T}.

\begin{figure*}
\centering
\includegraphics[width=1\textwidth]{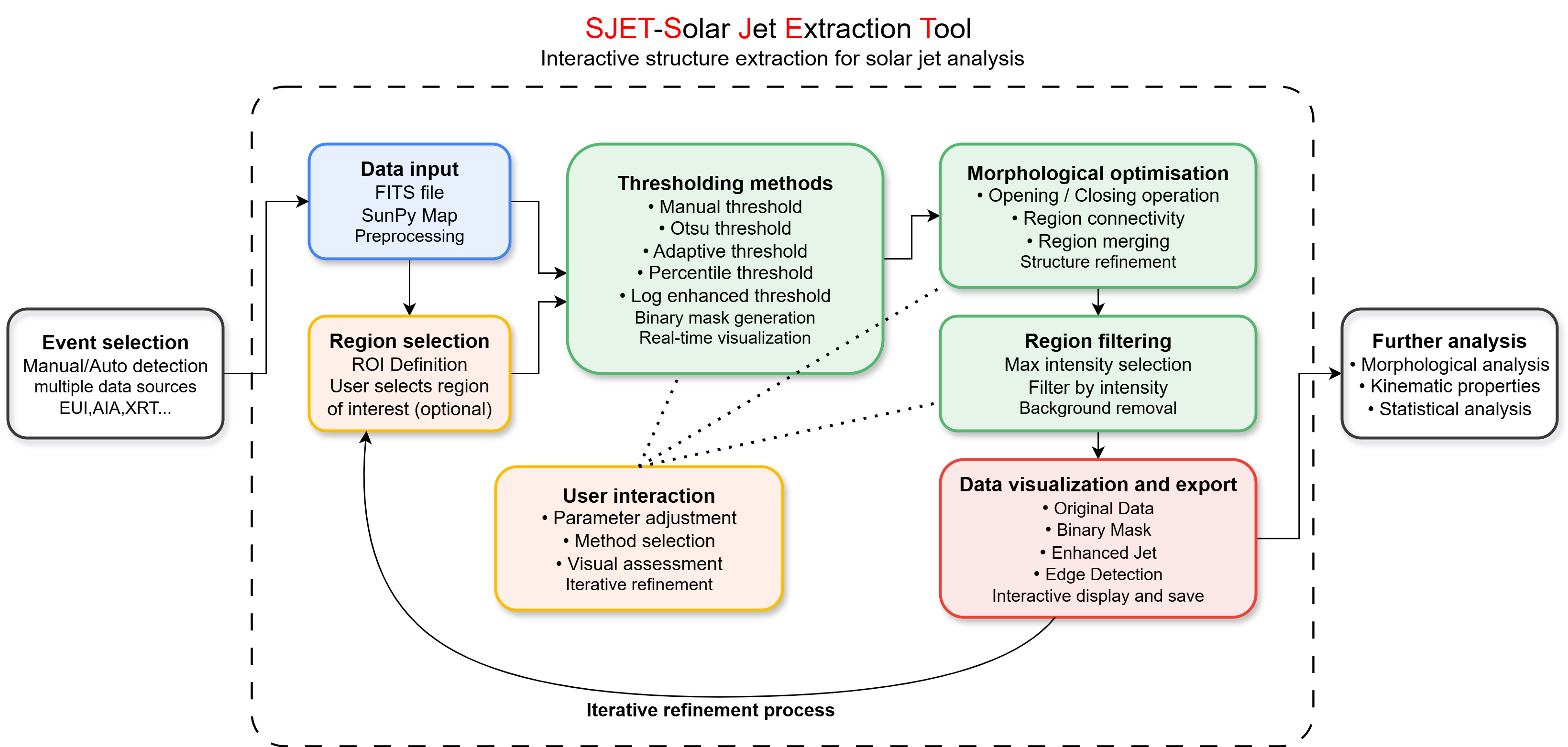}
\caption{SJET workflow illustrating the interactive jet extraction process. The workflow begins with event selection and data input through FITS files, followed by optional ROI definition. Users can apply one of five thresholding methods (manual, Otsu, adaptive, percentile, and log-enhanced) to generate binary masks, which then undergo morphological optimization and region filtering. The central user interaction component enables real-time parameter adjustment and method selection through an iterative feedback loop. Throughout the process, a four-panel visualization displays original data, binary masks, enhanced jet structures, and edge detection results. The workflow concludes with data export capabilities supporting further analysis. The dashed boundary encompasses the core SJET processing environment, emphasizing the tool's integrated and interactive design.}
\label{fig1}
\end{figure*}

Quantitative understanding of solar jet properties has been advanced through numerous large-sample statistical studies employing different identification methodologies. Early foundational work relied on manual visual identification, exemplified by \cite{1996PASJ...48..123S} who identified 100 X-ray jets from Yohkoh/SXT observations, establishing fundamental scaling relationships for jet lengths ($10^{4}$-$10^{5}$ km), velocities (10-1,000 km/s), and lifetimes (power-law distribution with index $\sim$1.2). Subsequent manual studies expanded across different solar regions and wavelengths: \cite{2015A&A...579A..96P} analyzed 18 polar jets using Hinode/XRT data, \cite{2016SoPh..291.1129N} compared network jets between coronal holes and quiet-Sun regions, \cite{2009SoPh..259...87N} examined 79 polar EUV jets using STEREO observations, and \cite{2016A&A...589A..79M} studied 20 active region jets across multiple AIA passbands. Larger manual surveys include \cite{2017ApJ...835...47K} who performed  a comprehensive statistical analysis of 301 macrospicular jets from  SDO/AIA 304 \AA\ observations over a 5.5-year period, investigating their temporal behavior and spatial properties in coronal holes and quiet Sun regions. Recent studies have also focused on the detailed investigation of solar jets using high-resolution space-based observations. In particular, \cite{2025A&A...702A.188N} conducted a recent analysis of 11 jets originating from a coronal bright point observed with the Solar Orbiter. As datasets grew larger, semi-automated approaches emerged, notably \cite{2023ApJS..266...17L} who developed the Semi-Automated Jet Identification Algorithm (SAJIA) to detect more than 1,200 off-limb jets during solar cycle 24, later expanded by \cite{2024ApJ...965...43S} to over 2,700 events. Building upon this foundation, \cite{2024ApJ...972..187L} developed the fully Automated Jet Identification Algorithm (AJIA) using U-NET neural networks, significantly improving precision from SAJIA's 0.34 2,010 true positives out of 7,890 candidates) to 0.81. Subsequently, progress has also been made in machine learning-based solar jet identification \citep{2025A&A...698A..50C}. Complementing these algorithmic approaches, citizen science methods have shown remarkable potential, with the Solar Jet Hunter project \citep{2024A&A...688A.127M} engaging volunteers to identify 883 jets from over 120,000 SDO/AIA images.

However, while these studies have established comprehensive solar jet event catalogs, the systematic extraction and quantitative analysis of jet morphological parameters across different identification methods remain significant challenges that require specialized tools and standardized workflows. Furthermore, the effectiveness of jet identification and extraction is strongly dependent on their location on the solar disk. Limb jets typically appear as bright linear structures extending outward from the chromosphere or transition region, with morphological features that are relatively easy to identify. In contrast, on-disk jets are particularly difficult to identify and extract due to projection effects and background contamination. This is why the vast majority of jets in the previously mentioned algorithms and citizen science methods are limb jets \citep{2023ApJS..266...17L,2024ApJ...965...43S,2024ApJ...972..187L,2024A&A...688A.127M,2025A&A...698A..50C}. The diversity of jets becomes particularly evident in large-sample event analysis and multi-instrument studies, motivating the development of standardized and reproducible analysis workflows for solar jet research.

In response to these challenges, this paper presents the Solar Jet Extraction Tool (SJET), an interactive tool for solar jet feature extraction, developed in Python. SJET adopts a visual interactive design philosophy, integrating multiple thresholding algorithms, morphological operations, and geometric parameter analysis into a unified workflow. The tool supports flexible region-of-interest (ROI) selection, real-time parameter adjustment, and immediate result feedback, effectively addressing the difficulties in parameter selection and result reproducibility encountered by traditional methods when processing complex jet morphologies. Through standardized processing workflows and comprehensive metadata preservation mechanisms, SJET provides a reliable technical foundation for quantitative analysis and large-sample statistical studies of solar jets.

This paper is organized into five sections. Section \ref{sec2} provides detailed descriptions of SJET's overall framework, core algorithms, and user interface design, including the technical implementation of five thresholding methods, morphological optimization strategies, and region merging algorithms. Section \ref{sec3} describes the selection criteria for test datasets and the mathematical principles for extracting geometric parameters of jets, with particular emphasis on the method for identifying start and end points based on circular regions and techniques for modeling the axis using Bézier curves. Section \ref{sec4} validates SJET's analytical capabilities through Solar Orbiter/EUI and SDO/AIA observational data and conducts quantitative comparisons with traditional slice analysis methods. Section \ref{sec5} summarizes SJET's main contributions, discusses current method limitations, and outlines future improvement directions and application prospects.

\begin{figure*}
\centering
\includegraphics[width=1\textwidth]{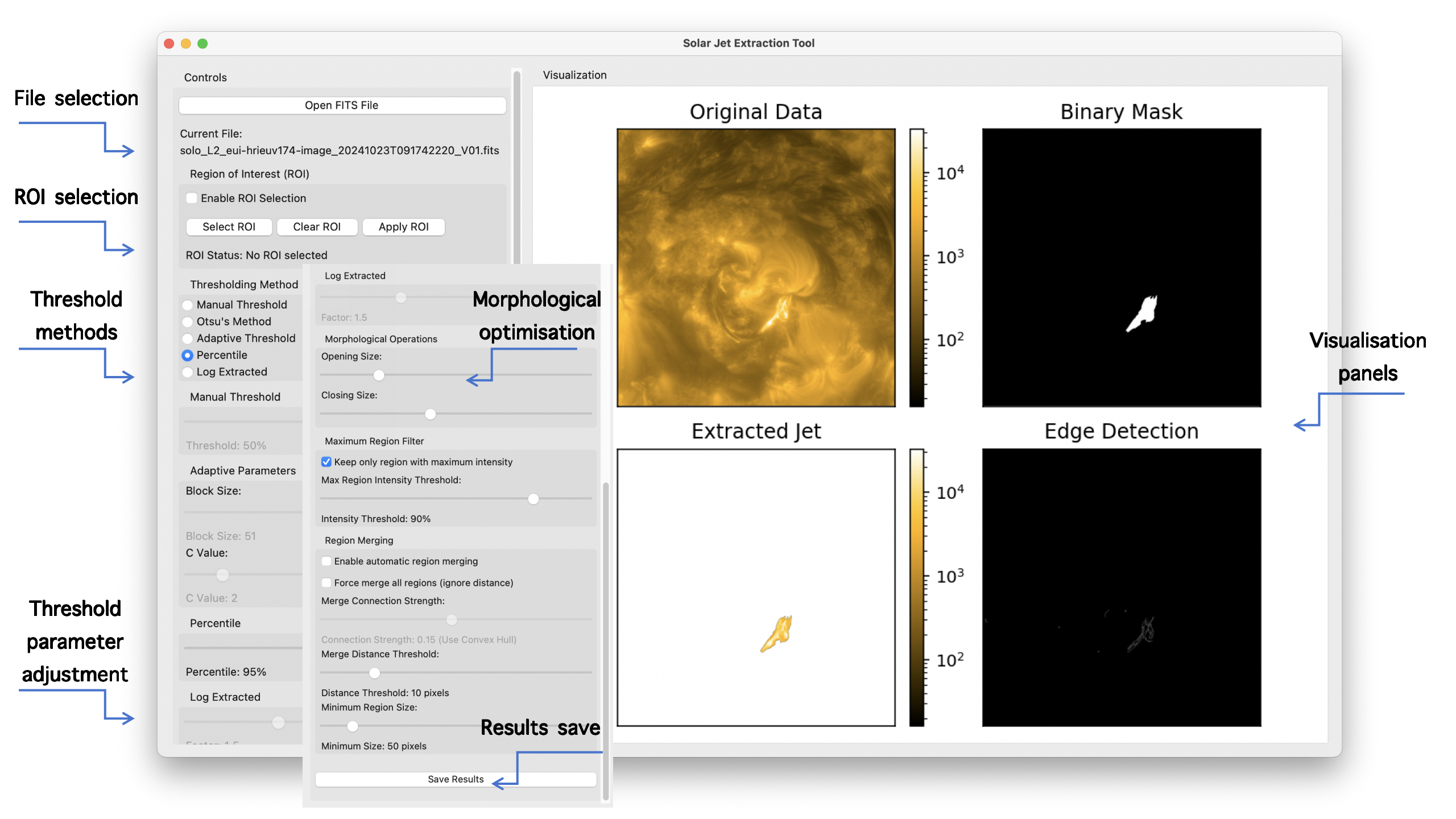}
\caption{SJET interactive interface and functionality demonstration. The left panel displays, from top to bottom: control panel, method selection, manual adjustment, and morphological optimization options. The right visualization panel presents an example of $\mathrm{HRI_{EUV}}$ analysis. When users achieve satisfactory results, they can click the save button in the lower left corner of the interface, which will save all FITS files and operation records.}
\label{fig2}
\end{figure*}

\section{SJET Design and Implementation}
\label{sec2}

\subsection{Overall Framework}

SJET's overall framework, shown in Fig. \ref{fig1}, follows a basic workflow of "input-processing-interaction-output." Based on identified solar jet events, users download and import FITS format data files representing characteristic moments of jets. The tool reads the data and performs necessary preprocessing. During the data preprocessing stage, the NaN values and standard fill values are replaced with zero. For the logarithmic enhancement thresholding method, negative pixel values are clamped to zero prior to the logarithmic transformation, as such values typically represent a negligible fraction of the image and do not introduce significant bias. Dynamic range normalization is applied for display purposes, while original physical units and coordinate metadata are preserved throughout. To improve processing efficiency and precision, SJET implements a flexible ROI selection functionality, allowing users to limit the analysis to a specific region of interest. This is especially valuable when analyzing complex active regions (e.g., regions containing other simultaneous activities). 

The implementation of SJET is built upon several open-source Python scientific computing libraries. Data reading and preprocessing are accomplished through the SunPy library \citep{sunpy_community2020,Mumford2020}, while image processing algorithms are primarily derived from scikit-image \citep{2014PeerJ...2..453V} and OpenCV \citep{bradski2000opencv}. Geometric computations rely on SciPy \citep{2020NatMe..17..261V} and NumPy \citep{2020Natur.585..357H}, and the user interface is constructed using Tkinter and Matplotlib \citep{2007CSE.....9...90H}. This integration of established scientific computing tools ensures both reliability and reproducibility of the analysis results.

\subsection{Thresholding Algorithms}
SJET's core functionality lies in the integration of five different thresholding algorithms, enabling users to make targeted algorithm selections based on specific observational data characteristics. The manual threshold method provides the most direct control, allowing users to adjust threshold levels through percentage sliders. This method offers simple implementation and demonstrates 
good segmentation performance when processing high-contrast jets, 
but requires careful tuning when the jet-to-background contrast is low 
or when background emission is spatially non-uniform. The Otsu automatic thresholding method, based on image histogram statistical properties, automatically determines optimal thresholds by maximizing inter-class variance, making it particularly 
suitable for solar images exhibiting bimodal intensity histograms, 
where bright jet structures are clearly separated from the darker 
background. For complex scenarios with non-uniform backgrounds, the adaptive thresholding method exhibits unique advantages. This algorithm considers local neighborhood features of each pixel, effectively handling non-uniform background issues in on-disk jets. The percentile thresholding method approaches the problem from a data statistical distribution perspective, allowing users to set thresholds based on intensity percentiles, which proves particularly effective when processing solar data 
with long-tail distribution characteristics, though the 
optimal percentile value is data-dependent and should be 
adjusted according to the brightness contrast of the specific jet. Targeting the large dynamic range characteristics of solar physics data, the logarithmic enhancement thresholding method enhances weak signal contrast through logarithmic transformation, achieving high detection precision in extraction of low-intensity jet structures.

\subsection{Morphological Processing and Region Analysis}
After obtaining binary masks processed by thresholding algorithms, SJET provides a series of morphological operations to optimize jet structure extraction results. Opening operations effectively remove small noise points and smooth jet boundaries, while closing operations fill holes within jets (resulting from specific thresholding algorithms or substructures inherent to jets) and connect adjacent structural fragments. The sizes of structural elements for these operations can be manually adjusted according to specific jet scales.

The systematic region merging functionality has been specifically designed for the physical characteristics of solar jets. Distance-based merging algorithms can integrate spatially proximate jet fragments into complete structures, while intensity-based filtering ensures retention of only regions with significant physical meaning. The "force merge" option can integrate all related regions into a single connected structure through convex hull algorithms, which is of great importance for studying jets with complex morphologies. Users should exercise caution when applying the force merge option to strongly curved or C-shaped jet structures: the convex hull operation fills the interior of the region, which can shift the geometric centre of the mask away from the true jet spine and lead to biased axis fitting. In such cases, manual adjustment of the Bézier control point (see Section~\ref{sec3}) is recommended.

\begin{figure}
\centering
\includegraphics[width=0.475\textwidth]{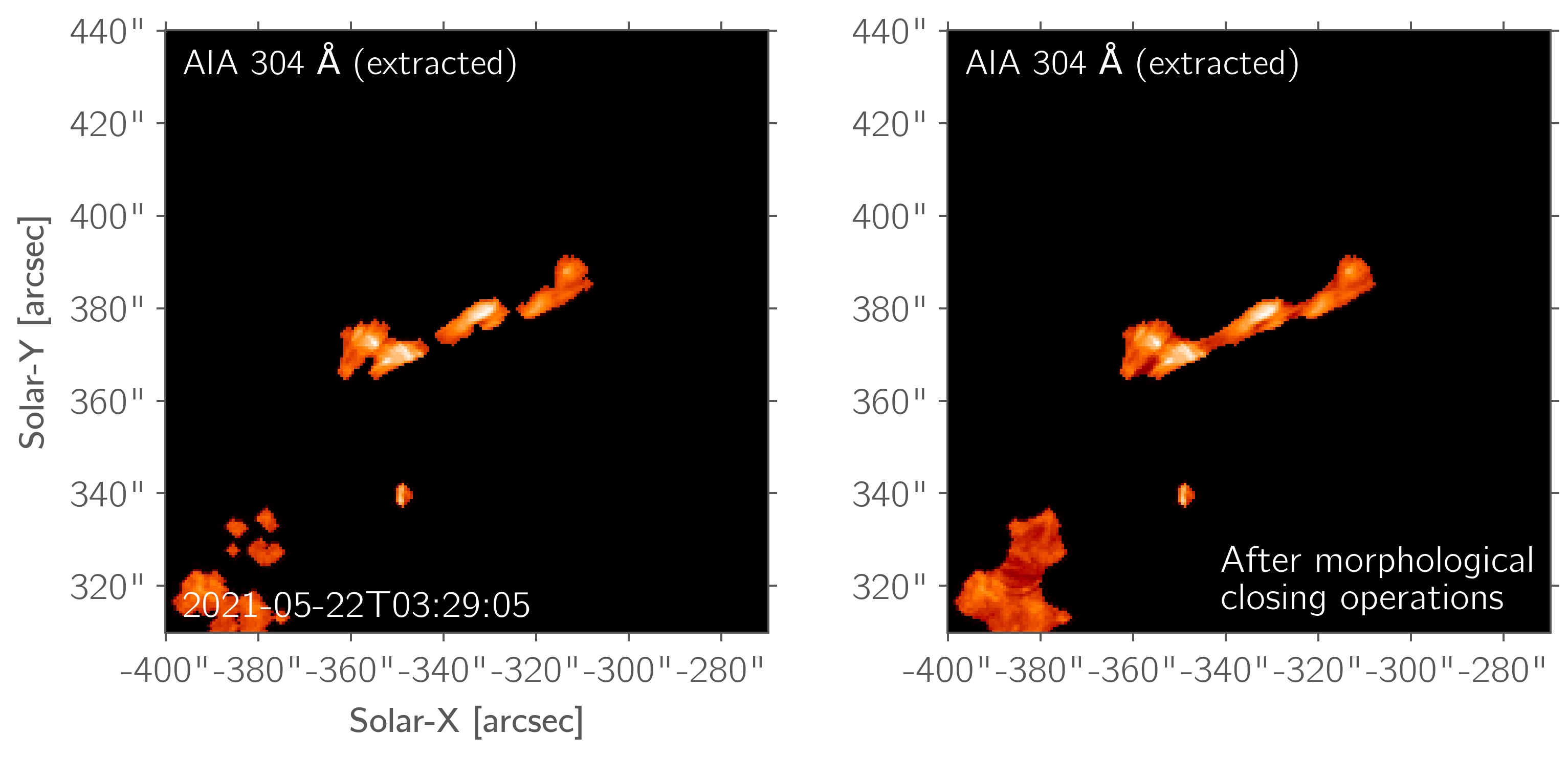}
\caption{Example of morphological processing. The panels depict an extracted AIA 304 \AA\ jet image and a further processed image (after morphological closing operations), respectively.}
\label{fig3}
\end{figure}

\begin{figure}
\centering
\includegraphics[width=0.475\textwidth]{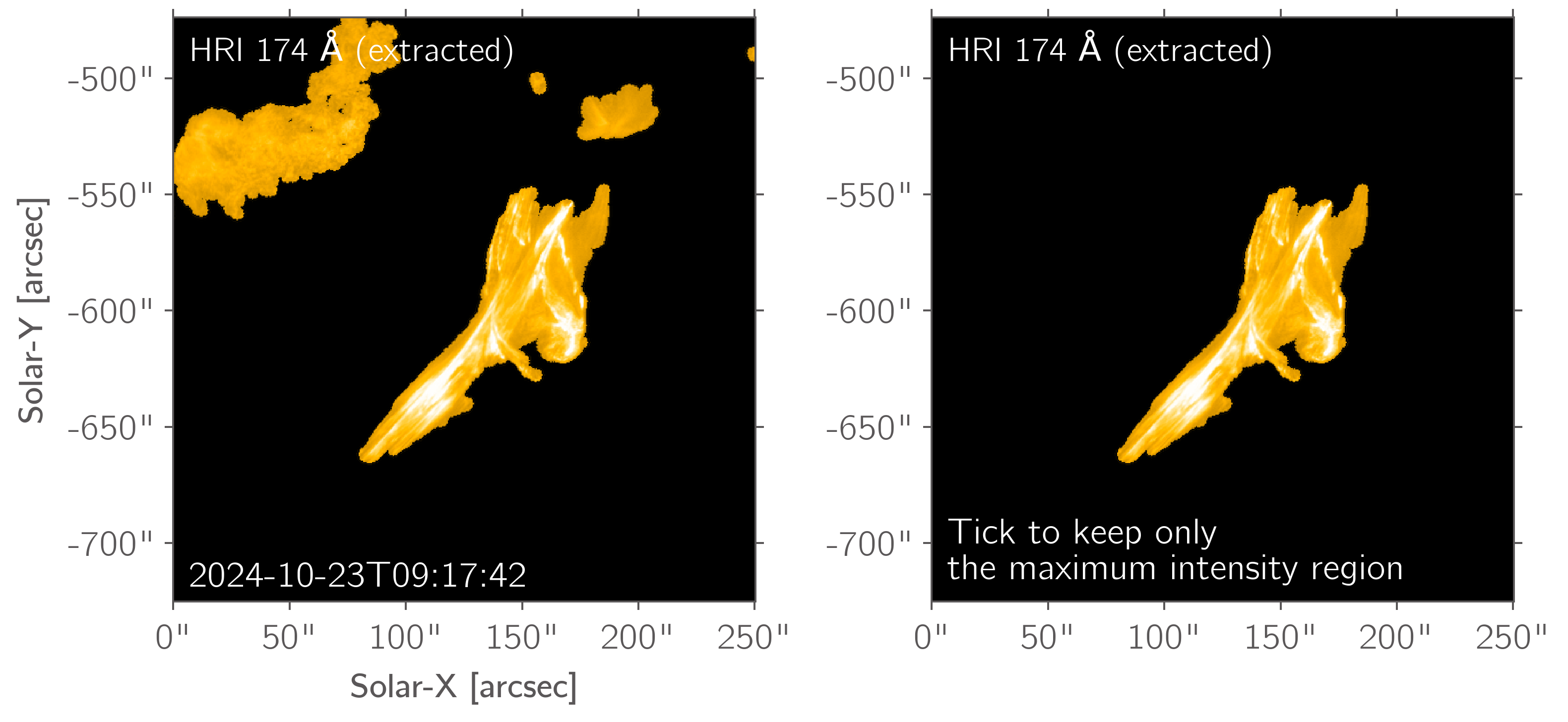}
\caption{Example of filtering regions by maximum intensity. The panels show an extracted EUI-$\mathrm{HRI_{EUV}}$ jet image and a further processed image (with a tick to keep only the region of maximum intensity), respectively.}
\label{fig4}
\end{figure}

In Figs. \ref{fig3} and \ref{fig4}, we test the morphological operations and the maximum intensity region filtering function, respectively.

\subsection{User Interface and Interactive Design}
SJET's graphical user interface adopts an intuitive layout design, with reasonable distribution of control panels and visualization areas. The control panel integrates all parameter adjustment functions, including threshold method selection, parameter sliders, and morphological operation controls. The visualization area employs a 2×2 subplot layout, simultaneously displaying original data, binary masks, extracted jet structures, and edge detection results, providing users with comprehensive feedback on the effects of processing.

All parameter adjustments are reflected in real-time visualizations, greatly improving parameter optimization efficiency. The system provides comprehensive result export functionality:

\textit{FITS format:} Binary masks, extracted jet intensity, edge detection results, and ROI masks (if applied), all preserving complete observational metadata.

\textit{PNG format:} High-resolution (300 DPI) visualizations including colorized intensity maps and grayscale masks.

\textit{ASCII text:} Processing parameters (ROI coordinates, merging settings, thresholding methods) recorded in companion files with standardized naming convention for full traceability.

This ensures reproducibility and facilitates integration with subsequent analysis pipelines.

\section{Dataset Selection and Geometric Parameter Extraction Algorithm}
\label{sec3}

\subsection{Observation Data}

To comprehensively validate SJET's analytical capabilities and scientific application value, we selected two representative types of solar observational data for analysis. We chose Solar Orbiter/EUI High Resolution Imager ($\mathrm{HRI_{EUV}}$, \cite{2020A&A...642A...8R}) 174 \AA\ observations as the latest test data, while utilizing SDO/AIA 304 \AA\ data as conventional reference data \citep{2012SoPh..275...17L}. The adoption of multi-instrument, multi-scale datasets enables comprehensive evaluation of SJET's performance under different observational conditions while establishing a data foundation for joint stereoscopic observations.

The $\mathrm{HRI_{EUV}}$ dataset consists of active region jet events from October 23, 2024. Solar Orbiter's orbital position at the time of observation (0.544 AU) enables $\mathrm{HRI_{EUV}}$ to observe solar atmosphere with 195 km/pixel image scale and 10-second temporal resolution. The selected time range spans 08:55 UT to 09:55 UT, targeting an active region near disk center.

The SDO/AIA 304 \AA\ dataset includes multiple jet activities from a single active region around May 22, 2021 \citep{2022FrASS...9.9799H,2023FrASS..1048467N}, focusing on jet events at 03:30 UT as the primary target, which exhibit multi-branch structures and irregular morphological features. Although AIA's 435 km/pixel image scale and 12-second temporal resolution are lower than $\mathrm{HRI_{EUV}}$, its full-disk field of view and continuous observations provide irreplaceable advantages for large statistical studies of jets.
Additionally, a supplementary AIA 304~\AA\ observation 
on September 27, 2021 at 11:42:53~UT is used to demonstrate SJET's built-in 
Gaussian FWHM width measurement (Section~\ref{sec4}).

\subsection{Jet Geometric Parameter Extraction and Temporal Evolution Analysis}

\begin{figure}
\centering
\includegraphics[width=0.475\textwidth]{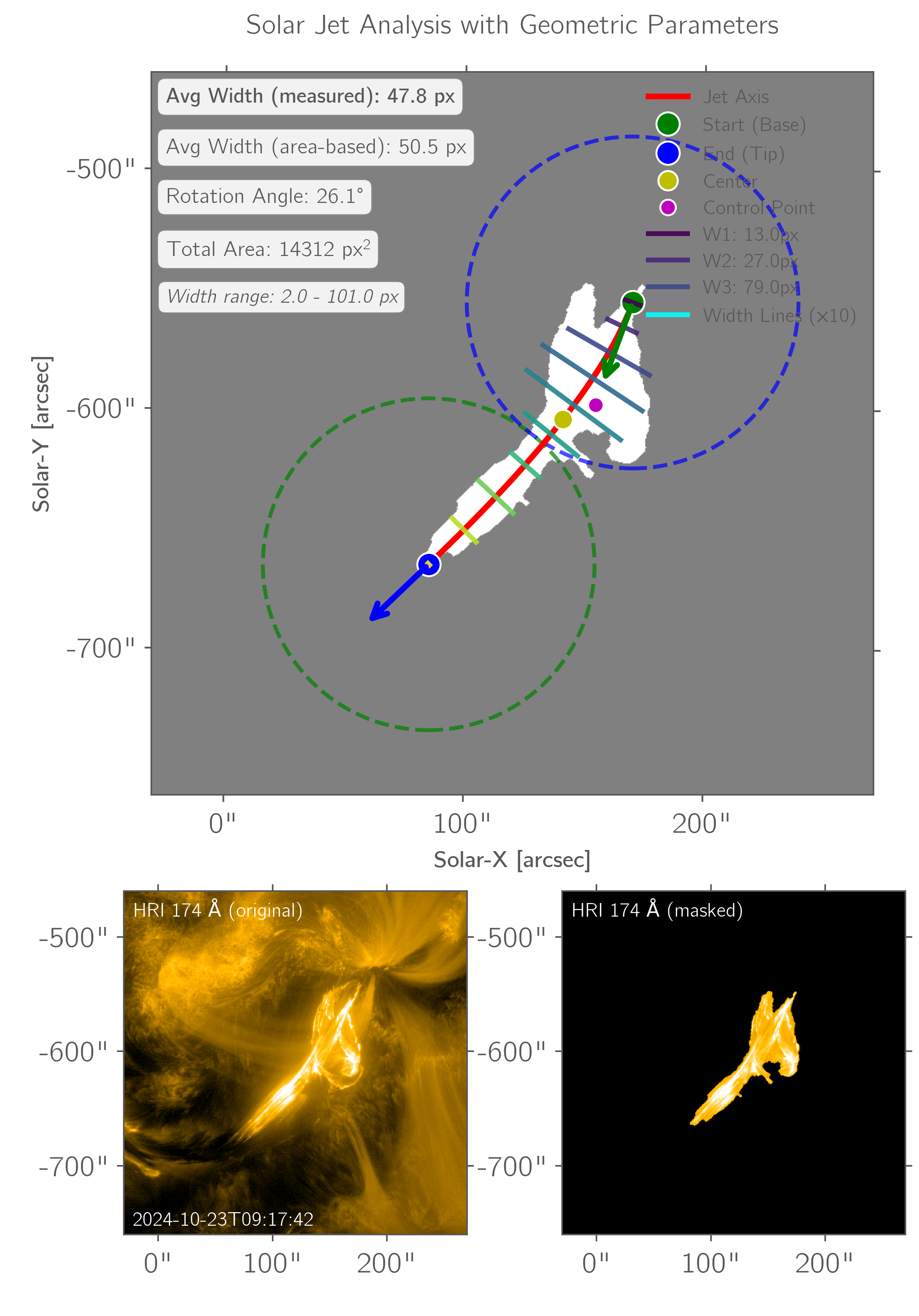}
\caption{Example of jet geometric parameter extraction. The upper panel displays the binary mask of the jet with annotations of the obtained geometric parameters and jet axis structures. The lower panels show the original $\mathrm{HRI_{EUV}}$ jet image and the extracted $\mathrm{HRI_{EUV}}$ jet image, respectively.}
\label{fig5}
\end{figure}

Building upon Section \ref{sec2}'s introduction to SJET's interface and processes for interactive extraction of jet structures, the geometric parameter extraction of jets based on binary masks constitutes the core analytical capability of the SJET tool (see Fig. \ref{fig5} for an example). The algorithm's first step involves identifying the two most distant pixels in the jet's binary mask, achieved through calculating Euclidean distances between all pixel pairs (for masks with areas exceeding 1000 pixels, an equidistant sampling method is activated to reduce computational load):
\begin{equation}
d_{\max} = \max_{i,j} \sqrt{(x_i - x_j)^2 + (y_i - y_j)^2}
\end{equation}
where $(x_i, y_i)$ and $(x_j, y_j)$ represent coordinates of any two pixels in the mask. This extremal point pair identification physically corresponds to determining the jet propagation direction, with the two extremal points located near the jet's source region and propagation front, respectively. After determining the extremal point pair, the algorithm constructs two circular counting regions centered on these points, with circular radius defined as:

\begin{equation}
r = \frac{d_{\max}}{2}
\end{equation}
When two equal-radius circles are centered on the most distant point pair, they produce overlap in the jet's central region, but each circle covers unequal areas of the jet. This inequality precisely reflects the asymmetric distribution characteristics of jet morphology. For standard jet morphologies, the jet width near the source 
region typically exceeds that at the propagation front 
\citep{1996PASJ...48..123S, 2009SoPh..259...87N,2016SSRv..201....1R, 2021RSPSA.47700217S}, 
although this assumption may not hold for blowout jets 
with broader spires \citep{2010ApJ...720..757M, 
2012ApJ...745..164S,2015Natur.523..437S, 2021ApJ...912L..15T}. The algorithm quantifies this width difference by counting mask pixels contained within each circular region:
\begin{equation}
N_i = |{(x,y) \in \text{mask} \cap C_i}|
\end{equation}
where $C_i$ represents the $i$-th circular region and $N_i$ denotes the number of pixels within that region. Based on physical intuition, regions containing more pixels correspond to wider jet bases, thus indicating jet source region locations. The start point discrimination criterion is expressed as:

\begin{equation}
\text{start point} = \arg\max_{P_i} N_i
\end{equation}

After determining the start and end point coordinates, the algorithm employs a quadratic Bézier curve to model the jet axis. Control point calculation is based on weighted combinations of geometric centers:
\begin{equation}
P_{\text{control}} = 2P_{\text{center}} - \frac{P_{\text{start}} + P_{\text{end}}}{2}
\end{equation}
where $P_{\text{center}}$ represents the geometric center coordinates of the jet mask, and $P_{\text{start}}$ and $P_{\text{end}}$ represent start and end point coordinates, respectively. This control point calculation formula ensures that the generated Bézier curve will always go through the center
coordinates of the mask. To address cases where the automatically computed control point deviates from the true jet spine — as can occur with strongly curved, C-shaped, or force-merged structures — SJET now provides a manual control point override. Users may specify an alternative control point coordinate, which replaces the geometric-centre-based estimate and allows the Bézier curve to be adjusted to better follow the physically meaningful jet axis. Both the automatic 
and manually adjusted control points are recorded in the output 
metadata for reproducibility. A demonstration of this 
feature, together with a sensitivity analysis of the Bézier axis 
extraction, is provided in Appendix~\ref{subsec:sensitivity}. Based on the parametric representation of the Bézier curve, we can calculate the actual length of the axis $L_c$ and define a curvature parameter:

\begin{equation}
\kappa = \frac{L_c - L}{L}
\end{equation}
where $L$ represents the straight-line distance from start to end points. Additionally, we can calculate the axis deflection angle from the jet base to its front, providing important geometric indicators for quantitatively describing the degree to which jet propagation paths deviate from straight lines, thereby offering insights into jet interactions with background magnetic fields.

\begin{figure*}
\centering
\includegraphics[width=1\textwidth]{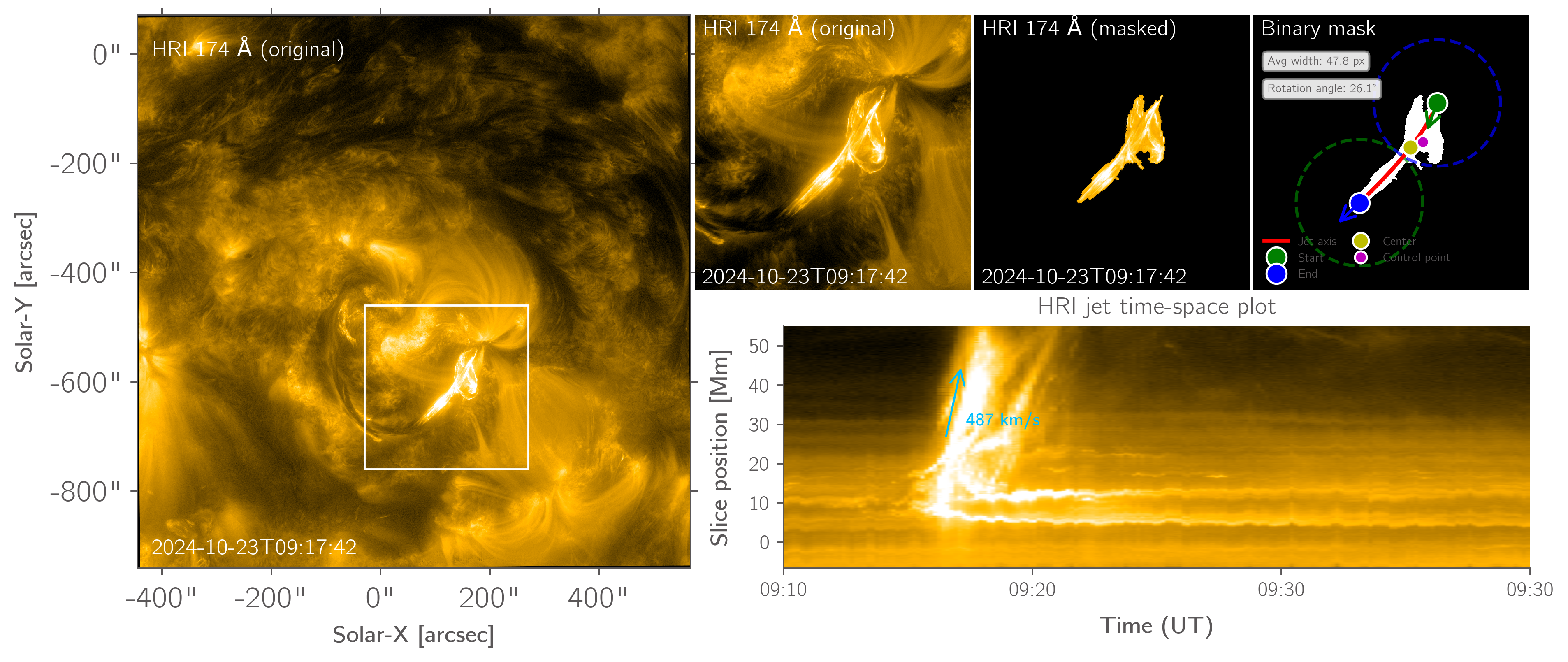}
\caption{SJET-based $\mathrm{HRI_{EUV}}$ jet analysis. The left panel presents the complete $\mathrm{HRI_{EUV}}$ image, while the right panels show the analysis results. Notably, the time-space plot, constructed using slices positioned along the fitted Bézier curve, reveals the jet propagation velocity through linear fitting.}
\label{fig6}
\end{figure*}

The width measurement process employs cross-sectional analysis techniques perpendicular to the axis. The algorithm selects ten measurement points at equal intervals along the Bézier curve, calculating the tangent direction at each measurement point to determine the perpendicular directions. Perpendicular width measurements are implemented through boundary detection algorithms, with local width $W_i$ defined as the maximum perpendicular jet structure extent. The average width is calculated through arithmetic averaging of all valid measurement points:

\begin{equation}
\bar{W} = \frac{1}{N} \sum_{i=1}^{N} W_i
\end{equation}
where $N$ represents the number of valid measurement points. This multi-point measurement strategy effectively suppresses the influence of local morphological irregularities on width estimation, providing more stable and reliable estimates of jet width. Additionally, we provide area-based jet mask width estimation 
methods as a complementary morphological parameter 
(alongside the Gaussian FWHM width described in 
Section~\ref{sec4}). Based on existing solar physics research experience, this paper proposes multiple geometric parameter analysis methods, providing an analytical framework for subsequent research that users can further develop according to their specific research needs.

\begin{figure*}
\centering
\includegraphics[width=1\textwidth]{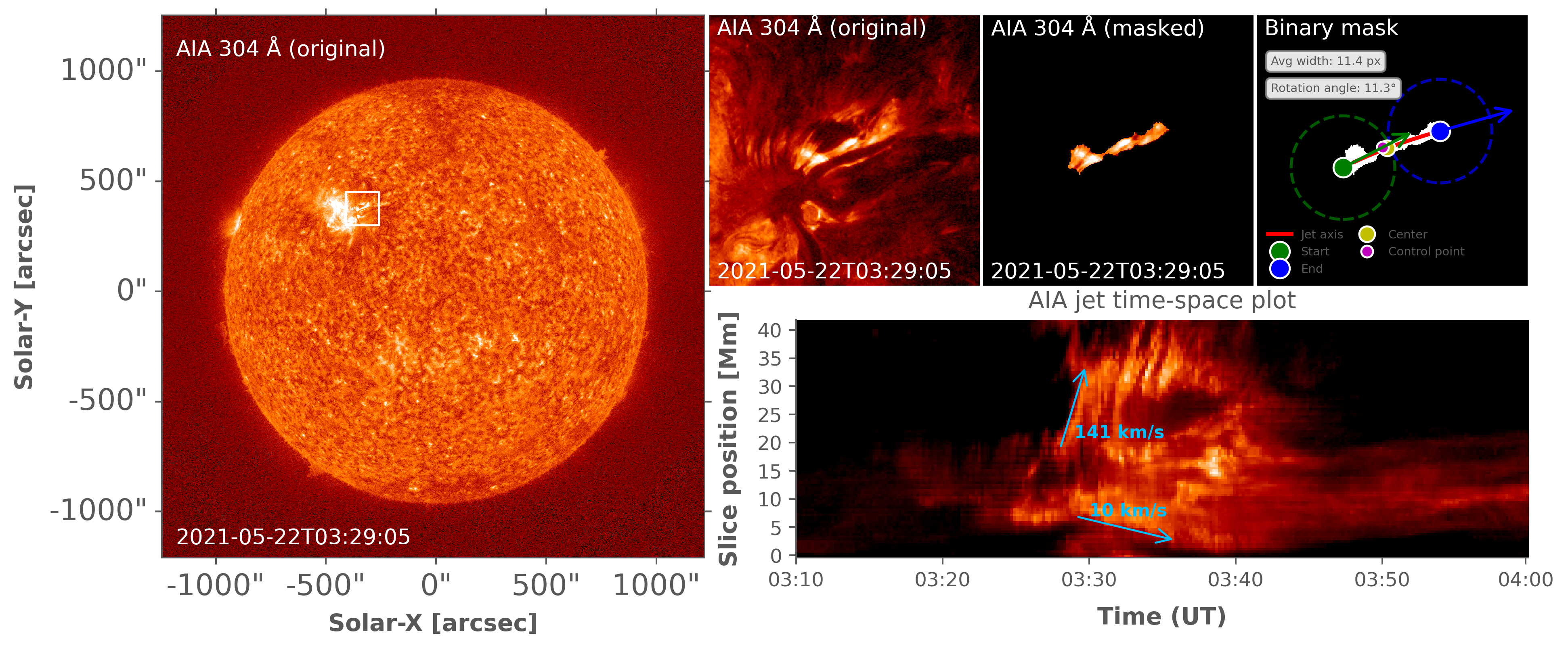}
\caption{SJET-based AIA 304 \AA\ jet analysis. Similar to Fig. \ref{fig6}, but the time-space plot simultaneously marks the velocities of bidirectional jet propagation.}
\label{fig7}
\end{figure*}

\section{Method Verification and Performance Assessment}
\label{sec4}

After introducing the test datasets and methods for mask-based jet analysis, this section presents datasets analysis results and compares SJET mask-based results with traditional analysis methods.

\subsection{SJET Analysis Results}
As shown in Fig. \ref{fig6}, the left subplot displays the complete $\mathrm{HRI_{EUV}}$ image, while the right subplot shows SJET analysis results. Figure \ref{fig5} displays the main geometric parameters of the extracted jet, including width (48.0 pixels) and rotation angle (26.1 degrees), and the generated time-space plot based on the fitted Bézier curve. Furthermore, we obtained the jet velocity (487 km/s) through simple linear fitting. Due to $\mathrm{HRI_{EUV}}$'s 16-bit dynamic range, its high dynamic range provide a good signal-to-noise ratio for SJET's threshold-based segmentation. Therefore, in actual extraction processes, we selected the log-threshold method (default parameters) and enabled the function to retain regions with maximum intensity. Analysis results demonstrate that despite coronal loop structures overlapping with jet bases in the line-of-sight direction, SJET is able to separate jet and coronal loop structures for this particular event. This indicates that for appropriate threshold parameters, threshold-based segmentation methods remain effective for complex on-disk structures.

Similar to the analysis using $\mathrm{HRI_{EUV}}$ data, Fig. \ref{fig7} displays an example using AIA 304 \AA\ images. We note that the AIA 304 \AA\ image contains relatively complex jet structures with fragmented jet morphologies. When  applying thresholding algorithms directly (percentile-based thresholding with parameters set to 98.8\%), complete connected jet structures cannot be obtained. For such complex jet structures, we further employed morphological optimization functions, connecting different jet components through closing operations (as shown in Fig. \ref{fig7}'s mask image). Based on this foundation, we obtained jet width (11.4 pixels) and rotation angle (11.3 degrees), and generated time-space plots of jets based on Bézier curves. For this complex jet, we observe that forward propagation is accompanied by numerous substructures, and the jet exhibits clear bidirectional propagation characteristics, with forward velocity (141 km/s) significantly exceeding backward velocity (10 km/s). This bi-directionally propagating on-disk jet is the so-called two-sided-loop jet which is important for diagnosing the local magnetic field structure \citep{1995Natur.375...42Y,2013ApJ...775..132J,2017ApJ...845...94T,2018ApJ...861..108Z,2019ApJ...883..104S,2019ApJ...887..220Y,2020MNRAS.498L.104W,2022MNRAS.516L..12T,2023MNRAS.520.3080T,2024ApJ...964....7Y,2026ApJ..1000....4Y}. It is worth noting that the feature analysed here corresponds to the jet base structure visible at 03:30 UT. The associated active region surge becomes more pronounced around 03:35 UT and propagates in a somewhat different direction. This highlights an inherent subjectivity in jet feature selection: the choice of which jet structure to analyse depends on the scientific question being addressed, and SJET does not automate this decision.

\subsection{Comparison with Traditional Methods}

The Gaussian slice fitting approach has been the established standard for solar jet width measurements since the foundational work of \cite{1996PASJ...48..123S}, with widespread adoption in subsequent studies \citep{2011ApJ...735L..43S,2012RAA....12..573C,2013ApJ...769..134M}. This methodology determines jet width by positioning measurement slices perpendicular to the jet axis and extracting the FWHM from a Gaussian fit to the resulting intensity profile. While robust and physically well-defined, its results depend on the number and placement of the manually selected slices.

\begin{figure*}
\centering
\includegraphics[width=1\textwidth]{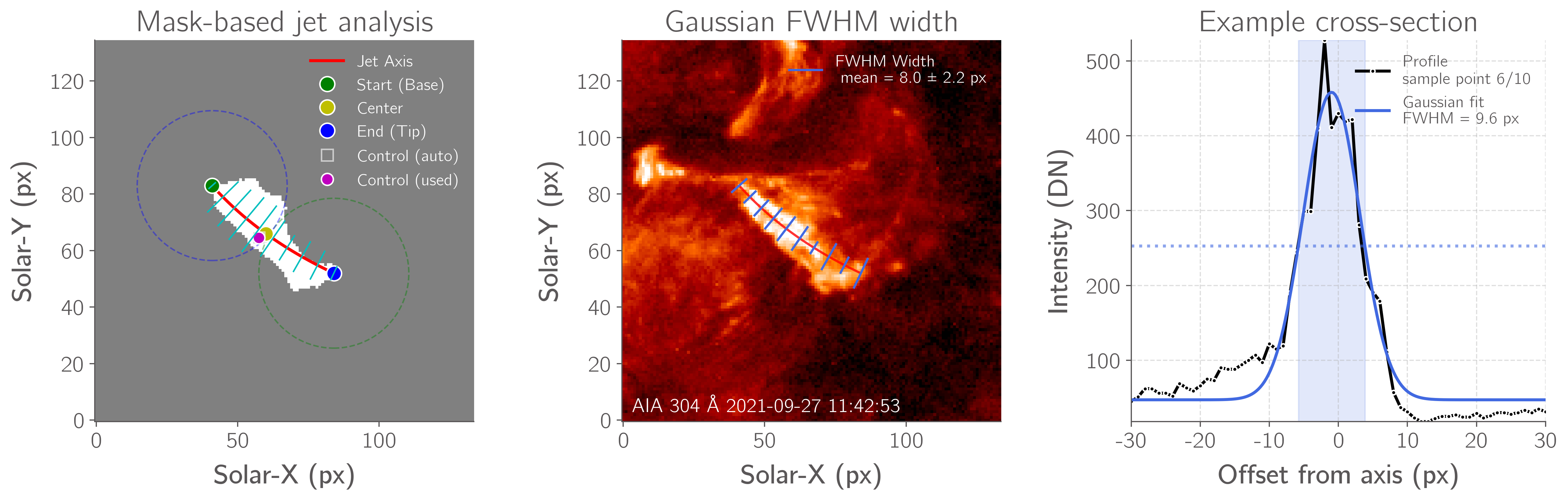}
\caption{Gaussian FWHM width measurement by SJET 
for a jet observed in the AIA 304~\AA\ channel 
(2021 September 27, 11:42:53~UT). 
Left: binary mask with the fitted Bézier axis (red) 
and mask boundary widths (cyan lines) at ten equally 
spaced locations along the axis. It is worth noting that, when no manual control points are entered, the control points (automatic) and control points (manual) coincide; please refer to the Appendix \ref{subsec:sensitivity} for further details.
Middle: original AIA 304~\AA\ image with the 
Gaussian FWHM widths (blue lines) overlaid at 
the same ten locations, yielding a mean FWHM of 
$8.00 \pm 2.17$~pixels. 
Right: example intensity profile (black) perpendicular 
to the jet axis at sample point 6 of 10, with the 
best-fit Gaussian (blue) and the corresponding 
FWHM interval (shaded region). }
\label{fig8}
\end{figure*}

SJET incorporates this approach as a built-in measurement: at each of the ten equally spaced 
locations along the fitted Bézier axis, a one-dimensional Gaussian profile — including a 
background baseline term — is fitted to the intensity distribution perpendicular to the axis 
on the original image, yielding a mean FWHM and its standard deviation across all valid 
cross-sections. To demonstrate this capability, we apply SJET to an AIA 304~\AA\ observation of a jet on 2021 September 27 at 11:42:53~UT (Fig.~\ref{fig8}), obtaining a mean FWHM of 
$8.00 \pm 2.17$~pixels and a mask boundary width of 12.70~pixels.

The ratio of FWHM to boundary width (63\%) is consistent with the physical expectation that 
the intensity core of the jet occupies a narrower region than its full detectable spatial extent. The two width parameters therefore carry complementary physical information: the FWHM 
reflects the bright intensity core of the jet and is directly comparable to traditional 
Gaussian slice measurements, while the boundary width captures the total morphological extent 
of the extracted structure. The standard deviation of the FWHM across the ten locations 
($\pm 2.17$~pixels) further quantifies the variation in jet cross-sectional width along 
its length — information that is not accessible from a single manually placed slice. Since the 
FWHM is derived from the original image intensity rather than the binary mask geometry, 
it is insensitive to the choice of thresholding method and morphological post-processing, 
making it a reproducible complement to the mask-derived width parameters. We therefore 
recommend that users report both measures according to their specific scientific objectives.

\section{Discussion and Conclusion}
\label{sec5}

We developed the SJET, implementing a complete workflow from interactive data processing to quantitative geometric parameter analysis. Through innovative technical integration and user-friendly interface design, the tool provides a standardized analysis platform for solar jet research. Validation analyses based on Solar Orbiter/EUI high-resolution 
observations and SDO/AIA data demonstrate SJET's ability to 
extract jet geometric parameters consistently across different 
observational conditions. However, we emphasise that manual 
intervention and user judgement remain central to the tool's 
operation: the selection of the jet event, the choice of 
thresholding method and parameters, and the definition of the 
ROI all require user expertise and introduce an inherent degree 
of subjectivity that the tool does not eliminate. In Appendix \ref{subsec:workflow}, we provide a comprehensive summary of the three-step workflow for the use of SJET, together with the jet parameters (\ref{tab:parameters}) derived from further analysis.

The fundamental challenges facing automated detection schemes in solar physics have been comprehensively demonstrated by \cite{2024ApJS..271....6R}, who assessed 14 widely applied automated coronal hole detection schemes using a community dataset of 29 SDO/AIA images. Their findings reveal that even for coronal holes—among the most visually distinct features in solar EUV images—the choice of automated detection scheme significantly affects derived physical properties. The substantial discrepancies between different automated schemes highlight fundamental challenges: noisy solar imagery, instrumental effects, and complex morphological variations make it extremely difficult to develop universally robust automated algorithms. If such difficulties arise for relatively obvious structures like coronal holes, the challenges become exponentially greater for solar jets, which exhibit far more complex morphologies, rapid temporal evolution, and subtle intensity contrasts against variable backgrounds.

The comparison of automated coronal hole detection algorithms provides valuable insights for SJET's semi-automated design philosophy. SJET's emphasis on user interaction allows researchers to guide extraction through real-time parameter adjustment and visual assessment. This approach utilizes human expertise to handle subtle morphological variations and complex background conditions that present difficulties for fully automated schemes. The algorithm for identifying start and end points based on circular regions  offers an objective method for determining jet propagation direction, enabling consistent geometric parameter extraction across different observational conditions through interactive parameter optimization. Further, the mask-based approach represents the jet as a two-dimensional structure rather than a one-dimensional linear feature, enabling richer morphological characterisation. We must emphasise that the reliability of the derived parameters depends on the quality of the extracted mask, which in turn is sensitive to the chosen thresholding configuration and morphological processing applied.
SJET's scientific value is reflected not only in precise analysis capabilities for individual jet events but more importantly in establishing a solid technical foundation for large-sample statistical studies. Through standardized processing workflows and comprehensive operation preservation mechanisms, SJET proposes a solution to the long-standing issues in solar jet research regarding method inconsistency and result comparison difficulties.

SJET's application prospects in the Comprehensive Solar Energetic Electron event Catalogue (CoSEE-Cat, \cite{2025A&A...701A..20W}) merit particular attention. CoSEE-Cat integrates observational data from multiple instruments on Solar Orbiter, providing valuable data resources for analysing solar energetic electron events\footnote{\url{https://coseecat.aip.de/}}. By applying SJET to large-sample jet analysis within the CoSEE-Cat event database, standardized extraction and analysis workflow can establish statistical correlations between jet parameters and the characteristics of released energetic electrons, holding important scientific value for understanding solar energetic particle acceleration mechanisms.

With the successive deployment of Solar Orbiter and future next-generation observational assets, solar physics faces unprecedented challenges regarding the analysis of high-resolution and diverse data. In this context, numerous Python-based packages for solar physics have recently emerged, providing scientists with a new platform for analysing and processing data with greater freedom and efficiency \citep{2020ApJ...890...68S,2020JOSS....5.2732S,2020zndo...4274931B,2022FrASS...9.4137K,2023SoPh..298...36N,2025RAA....25k5015Y,2025RASTI...4af053N}\footnote{\url{https://heliopython.org/}}. We also note that complementary jet catalogues compiled 
from SDO/AIA observations are available to the 
community (e.g., the catalogue maintained by 
A.~Shendrik at the St.~Petersburg Branch of SAO~RAS%
\footnote{\url{https://solar.sao.ru/coronal-jets-catalog/}}), 
which may serve as valuable benchmark datasets for 
future validation and large-sample applications of SJET. SJET's development experience demonstrates that integrating multiple algorithms into interactive platforms can effectively balance advantages of automated processing and human judgment, providing a design approach with reference value for processing other types of solar activity phenomena. Although current tool still has space for improvement, its exploration in establishing standardized workflows and preparing for large-sample analysis provides beneficial technical references for further development of solar jet research.

\section*{Acknowledgements}
We would like to thank the reviewers for their constructive comments, which have significantly improved the readability of our paper and the rigour of the tool. S.T. thanks Daniel El\'ias N\'obrega-Siverio 
for drawing  attention to the coronal jet 
catalogue maintained by 
A.~Shendrik. We also would like to acknowledge the SDO science team for providing the data. Solar Orbiter is a space mission of international collaboration between ESA and NASA, operated by ESA. The EUI instrument was built by CSL, IAS, MPS, MSSL/UCL, PMOD/WRC, ROB, LCF/IO with funding from the Belgian Federal Science Policy Office (BELPSO); the Centre National d'Etudes Spatiales (CNES); the UK Space Agency (UKSA); the Bundesministerium f\"ur Wirtschaft und Energie (BMWi) through the Deutsches Zentrum f\"ur Luft- und Raumfahrt (DLR); and the Swiss Space Office (SSO). The AIP team was supported by the German Space Agency (DLR), grant number \mbox{50 OT 2304}. A.W. and J.M. also acknowledge funding by the European Union’s Horizon Europe research and innovation program under grant agreement No. 101134999 (SOLER). Y.S. was supported by the Natural Science Foundation of China (12573059, 12173083).

\section*{Data Availability}
To support reproducible research and facilitate community adoption, the complete SJET source code will be made publicly available through a dedicated GitHub repository \footnote{\url{https://github.com/songsolarphysics/SJET}} upon publication. This repository will include comprehensive installation documentation, usage examples, and the complete codebase with detailed comments explaining the implementation of all algorithms described in this paper. Additionally, we will provide representative test datasets directly through the repository, including carefully selected examples from both Solar Orbiter/EUI $\mathrm{HRI_{EUV}}$ and SDO/AIA 304 \AA\ observations that demonstrate different jet morphologies and complexity levels. These test datasets will be accompanied by analysis parameter files that reproduce the results presented in Sect.~\ref{sec4}, enabling new users to immediately validate their installation and familiarize themselves with SJET's capabilities.

For accessing the complete observational datasets used in this study, researchers should follow established mission data access procedures. Solar Orbiter/EUI observations can be obtained through the Solar Orbiter Archive (SOAR) at https://soar.esac.esa.int/soar/, which provides comprehensive access to all instrument data with appropriate calibration information. In particular, EUI-HRI data are obtained from the latest data release 6.0 \citep{euidatarelease6}. SDO/AIA data remains freely accessible through the Joint Science Operations Center (JSOC) at http://jsoc.stanford.edu/ or alternatively through the Virtual Solar Observatory (VSO) at https://sdac.virtualsolar.org/, both offering flexible query capabilities for archived observations.

\section*{Conflict of Interest}
The authors declare no conflicts of interest.



\bibliographystyle{mnras}
\bibliography{ref} 




\begin{appendix}

\section{SJET Analysis Workflow}
\label{subsec:workflow}

\begin{table}
\centering
\caption{Solar jet parameters extracted by SJET. Geometric parameters are extracted directly from the binary mask. The boundary-based width $\bar{W}$ is obtained through perpendicular boundary detection at $N$ sample points along the jet axis, while the area-based width $W_{\rm area}$ is calculated from the total mask area $A$ divided by the curve length. The rotation angle $\theta_{\rm def}$ measures the angular change from jet base to tip. Kinematic analysis parameters are derived from intensity profiles along the extracted jet axis across multiple frames. For bidirectional jets (e.g., two-sided-loop jets), forward and backward velocities are determined separately through linear fitting of the corresponding propagation fronts in the time-space plot.}
\label{tab:parameters}
\begin{tabular}{llll}
\hline
\hline
Parameter & Method & Expression  \\
\hline
\multicolumn{4}{c}{\textit{Geometric and Morphological Parameters}} \\
\hline
\multirow{3}{*}{Jet width} & Boundary-based & $\bar{W}$ (mean of $W_i$)  \\
 & Area-based & $W_{\rm area} = A / L_c$  \\

& Gaussian FWHM & $\bar{W}_{\rm FWHM}$ \\
\hline
Jet length & Curve length & $L_c$  \\
\hline
Rotation angle & Deflection & $\theta_{\rm def}$  \\
\hline
Source region & Start point & $(x_{\rm start}, y_{\rm start})$ \\
\hline
\multicolumn{4}{c}{\textit{Kinematic Analysis Parameters}} \\
\hline
Onset time & Start time & $t_0$  \\
\hline
Propagation & Bidirectional & forward/backward  \\
\hline
\multirow{2}{*}{Velocity} & Forward & $v_{\rm f} = \Delta d_{\rm f} / \Delta t$ \\
 & Backward & $v_{\rm b} = \Delta d_{\rm b} / \Delta t$ \\
\hline
\end{tabular}
\end{table}

The complete SJET analysis workflow consists of three main steps:

\begin{enumerate}
    \item \textbf{Event Identification and Data Selection}: Identify solar jet events from observational data and determine the characteristic time range and key frames that capture the jet evolution. Users download and prepare FITS format data files representing these critical moments.
    
    \item \textbf{Interactive Jet Mask Extraction}: Load the selected FITS files into SJET and extract jet binary masks through the interactive interface. This process involves:
    \begin{itemize}
        \item Optional ROI selection to focus on the jet region
        \item Thresholding method selection (manual, Otsu, adaptive, percentile, or log-enhanced)
        \item Real-time parameter adjustment with visual feedback
        \item Morphological optimization (opening/closing operations)
        \item Region filtering and merging for complex jet structures
    \end{itemize}
    The tool provides a four-panel visualization (original data, binary mask, extracted jet, edge detection) for immediate quality assessment. All processed results, including binary masks and metadata, are exported in FITS format with complete operation records for reproducibility.
    
    \item \textbf{Geometric Parameter Extraction and time-space Analysis}: Apply the geometric parameter extraction algorithm to the optimized binary masks to automatically compute jet properties (Table \ref{tab:parameters}). For kinematic analysis, intensity profiles are extracted along the identified jet axis across all frames to construct time-space diagrams, from which propagation velocities and onset times are determined through linear fitting.
\end{enumerate}

For large mask regions exceeding 1,000 pixels, SJET employs a random sampling strategy when identifying extremal point pairs, reducing computational complexity from $O(n^2)$ to $O(n)$ while maintaining identification accuracy.

When multiple jet structures are present within the same field of view, the analysis is performed sequentially on an event-by-event basis. After completing the extraction and saving the parameters for one jet, the user can immediately initiate a new ROI selection within the same FITS image to isolate and process additional jet structures, without the need to reload the data.

\section{Sensitivity of Bézier Axis Extraction and 
Manual Control Point Override}
\label{subsec:sensitivity}

\begin{figure*}
\centering
\includegraphics[width=1\textwidth]{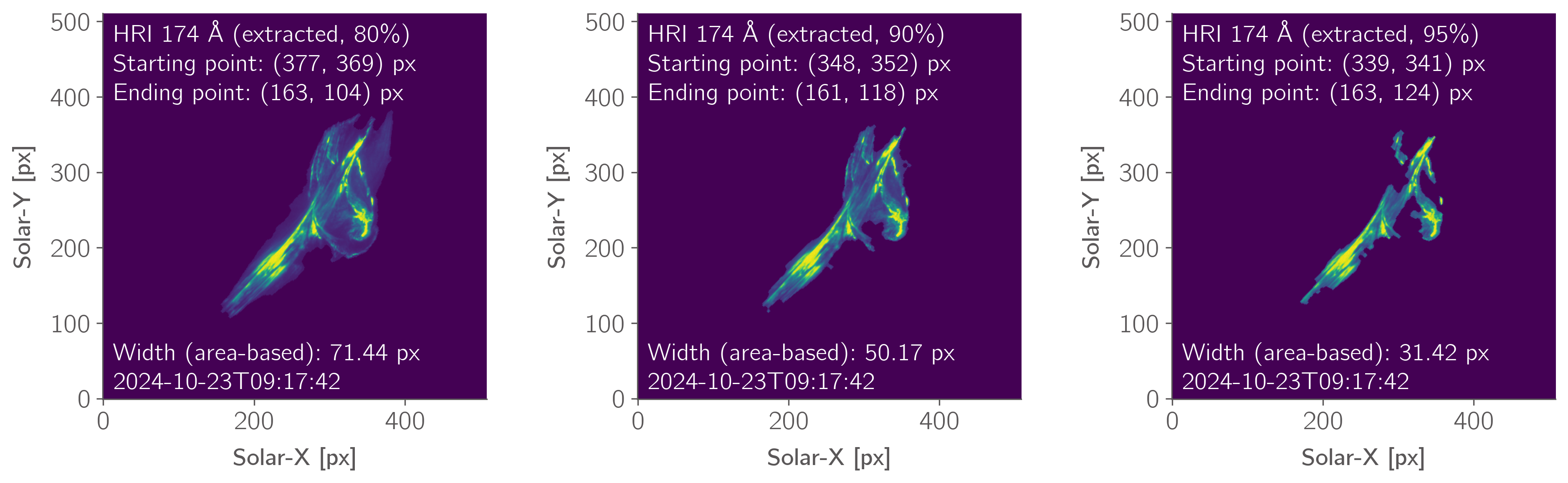}
\caption{Sensitivity test of SJET's Bézier axis extraction 
to threshold variation, applied to the same 
$\mathrm{HRI_{EUV}}$ jet at three percentile threshold 
levels (left: 80\%, centre: 90\%, right: 95\%). 
Each panel shows the extracted jet structure with the 
identified start point, end point, and area-based width 
annotated. As the threshold increases, the mask boundary 
progressively tightens: the area-based width decreases 
from 71.44~pixels at 80\% to 50.17~pixels at 90\% and 
31.42~pixels at 95\%, while the start and end points 
shift inward by $\sim$38 and $\sim$20~pixels respectively 
between the 80\% and 95\% cases.}
\label{fig:sensitivity}
\end{figure*}

To quantitatively assess how sensitive the derived 
geometric parameters are to variations in the extraction 
threshold, we applied three percentile threshold levels 
— 80\%, 90\%, and 95\% — to the same 
$\mathrm{HRI_{EUV}}$ jet (Fig.~\ref{fig:sensitivity}).

The identified endpoints remain spatially stable across 
all three threshold levels. The start point shifts by 
only $\sim$38~pixels between the 80\% and 95\% cases 
(from (377,~369) to (339,~341)), and the end point 
shifts by $\sim$20~pixels (from (163,~104) to 
(163,~124)). This stability reflects the fact that the 
extremal point pair identification is based on the 
global geometry of the mask rather than its precise 
boundary, making it relatively robust to moderate 
threshold variations.

The area-based width shows a stronger sensitivity 
across the three levels (71.44, 50.17, and 
31.42~pixels), as it integrates over the total mask 
area and is therefore more susceptible to the 
inclusion or exclusion of diffuse peripheral emission. 
As a practical consistency check, we recommend 
comparing the area-based width with the Gaussian FWHM 
width derived from the original image intensity: a 
large discrepancy between the two may indicate that 
the mask either over-includes diffuse non-jet emission 
at a loose threshold or under-extracts the jet core 
at a tight threshold, and the threshold should be 
adjusted accordingly.

\begin{figure*}
\centering
\includegraphics[width=1\textwidth]{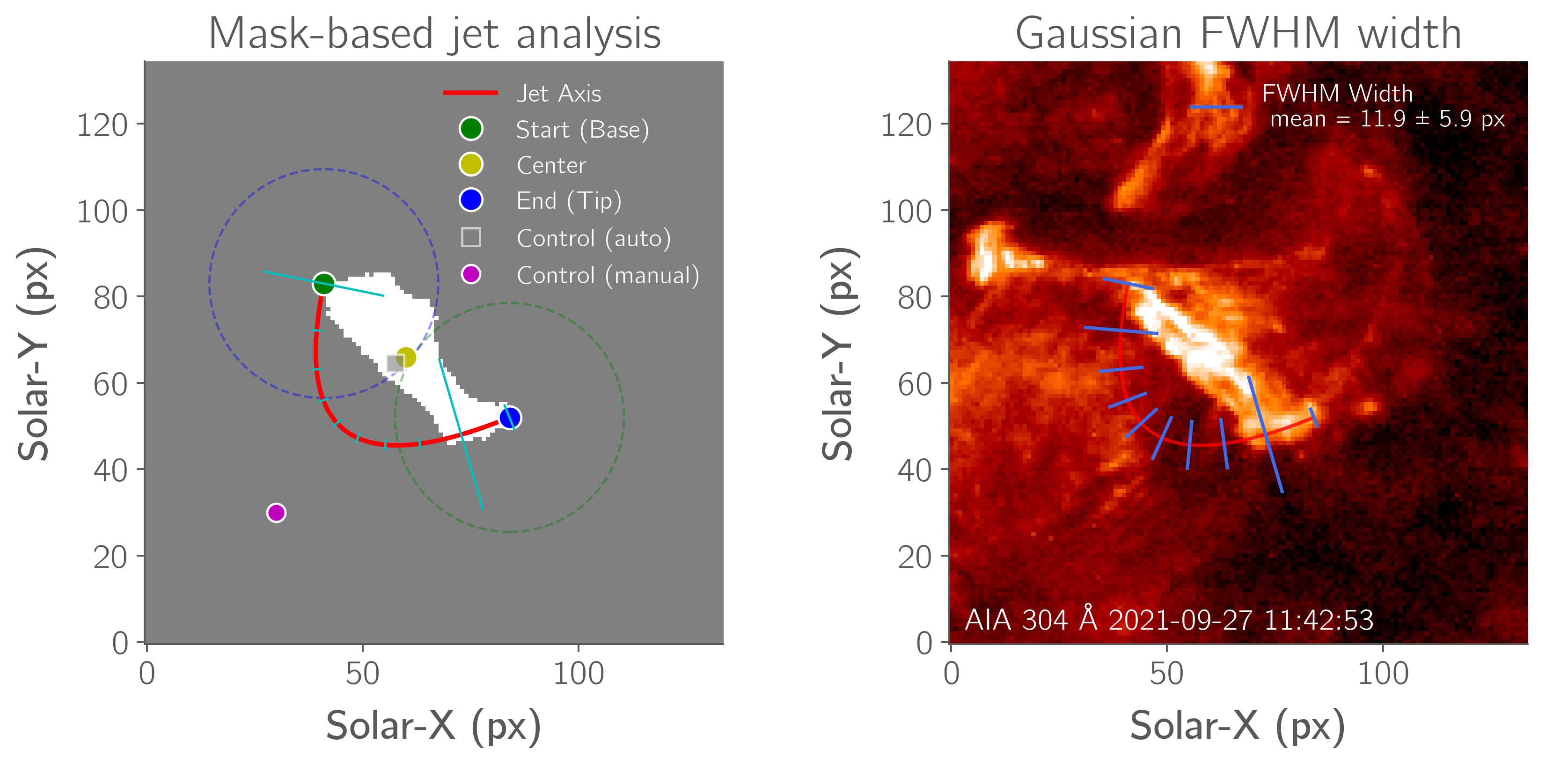}
\caption{Demonstration of SJET's manual control point 
override and Gaussian FWHM width measurement for an 
AIA 304~\AA\ jet (2021 September 27, 11:42:53~UT). 
Left: binary mask with the fitted Bézier axis (red 
solid curve), identified start point (green), 
geometric centre (yellow), end point (blue), 
automatically computed control point (grey square), 
and a deliberately misplaced manual control point 
(magenta circle at (30,~30)), which lies well outside 
the jet structure. The resulting Bézier curve visibly 
deviates from the jet spine, illustrating how an 
incorrect control point placement distorts the axis 
fit. In practice, users should adjust the control 
point to lie along the jet spine to obtain a 
physically meaningful axis. Right: original AIA 
304~\AA\ image with the Gaussian FWHM widths 
(blue lines) overlaid at ten equally spaced locations 
along the fitted axis, yielding a mean FWHM of 
$11.9 \pm 5.9$~pixels. The relatively large standard 
deviation reflects the axis distortion introduced 
by the misplaced control point.}
\label{fig:gaussain_test}
\end{figure*}

Fig.~\ref{fig:gaussain_test} demonstrates the manual control 
point override feature using an AIA 304~\AA\ jet image. 
In this example, the manual control point is 
deliberately placed at (30,~30), well outside the 
jet structure, to illustrate the consequence of 
an incorrect control point placement. The resulting 
Bézier axis (red solid curve in the left panel) 
curves strongly away from the true jet spine. 
The impact on the derived parameters is clearly 
visible in the right panel: the Gaussian FWHM 
measurements along the distorted axis yield a 
mean of $11.9 \pm 5.9$~pixels, with a 
standard deviation that is large relative to 
the mean, indicating that the cross-sections 
are sampling inconsistent regions of the image 
rather than the jet core.

This example highlights two practical points. 
First, when the automatic control point — based 
on the geometric centre of the mask — deviates 
significantly from the jet spine, as can occur 
for curved, C-shaped, or force-merged structures, 
the manual override should be used to correct 
the axis. Second, the standard deviation of 
the Gaussian FWHM across the ten measurement 
locations serves as a useful diagnostic: a 
large relative standard deviation may indicate 
that the fitted axis does not follow the jet 
spine consistently, and that the control point 
should be adjusted.

\end{appendix}

\bsp	
\label{lastpage}
\end{document}